\documentclass[aps,nofootinbib,longbibliography,prd,twocolumn]{revtex4-1}
\usepackage{graphicx}
\usepackage{amsmath,amssymb}
\usepackage[colorlinks,citecolor=blue,linkcolor=blue,urlcolor=blue]{hyperref}
\usepackage{mathrsfs}

\begin{document}

\title{Small black/white hole stability and dark matter}

\author{Carlo Rovelli${}^1$  and Francesca Vidotto${}^{2}$\\[-2mm]{\ }}

\affiliation{$^{1}$  CPT, Aix-Marseille Universit\'e, Universit\'e de Toulon, CNRS, Marseille, France;\\
$^{2}$ University\,of\,the\,Basque\,Country UPV/EHU, Departamento\,de\,F\'isica\,Te\'orica, 
E-48940 Leioa, Spain.}

\date{\small\today}

\begin{abstract}
\noindent We show that the expected lifetime of white holes formed as remnants of evaporated black holes is consistent with their production at reheating. We give a simple quantum description of these objects and argue that a quantum superposition of black and white holes with large interiors is stable, because it is protected by the existence of a minimal eigenvalue of the area, predicted by Loop Quantum Gravity. These two results support the hypothesis that a component of dark matter could be formed by small black hole remnants.
\end{abstract}

\maketitle


\section{Remnants}

The~possibility~that~remnants~of~evaporated~black holes
form~a~component~of~dark matter~was~suggested by
MacGibbon~\cite{J.H.MacGibbon1987}
~thirty~years~ago
and has~been~explored by~many~authors
\cite{Barrow1992,Carr:1994ar,Liddle1997,Alexeyev2002,Chen2003,Barrau:2003xp,Chen2004,Nozari2008}. 
There~are~no~strong observational constraints
on~this~possible~contribution to~dark matter 
\cite{Carr2016}; 
the~weak~point~of~this~scenario~has~been, so~far, the obscurity~of~the~physical~nature~of~the~remnants. 

The situation has changed recently because of the 
realisation that conventional physics provides a candidate 
for remnants: small-mass white holes with a large interiors \cite{Rovelli2014h,Haggard2014,DeLorenzo2016}. 
In addition, quantum gravity indicates that these \emph{are}
indeed produced at the end of the evaporation \cite{Christodoulou2016,Christodoulou2018,Bianchi2018,DAmbrosio2018,Rovelli2018a}. 
Here we show that the remnant lifetime predicted in \cite{Bianchi2018} 
is remarkably consistent with the production of primordial black holes 
at the end of inflation. 
More precisely, the rather strict constraints that 
the model sets on the time scales of the lifetime of black and white holes 
happen to match with a cosmological window where primordial black hole 
production is expected. 
A~preliminary version of this result was posted in 
\cite{Rovellia}.

Open questions are the stability and the eventual quantum properties 
of these remnants. 
It was suggested in \cite{Bianchi2018} that these 
remnants may be stable because quantum gravity dumps the Planck 
scale perturbations required to trigger their instability. 
Here we analyse the situation a bit more in detail by studying the stability 
of the remnants using a simple quantum model that captures 
the dynamical processes black and white holes can undergo. 
The model indicates that a quantum superposition of Planck size white and black holes should be stable, because of the large interior volume and the area gap, i.e. the presence of a minimal non-vanishing eigenvalue in the area spectrum according to 
Loop Quantum Gravity. 

These two results support the hypothesis that a component of dark matter could be formed by small black hole remnants. 

\begin{figure}[b]\label{Kruskal}
\includegraphics[width = .3 \textwidth]{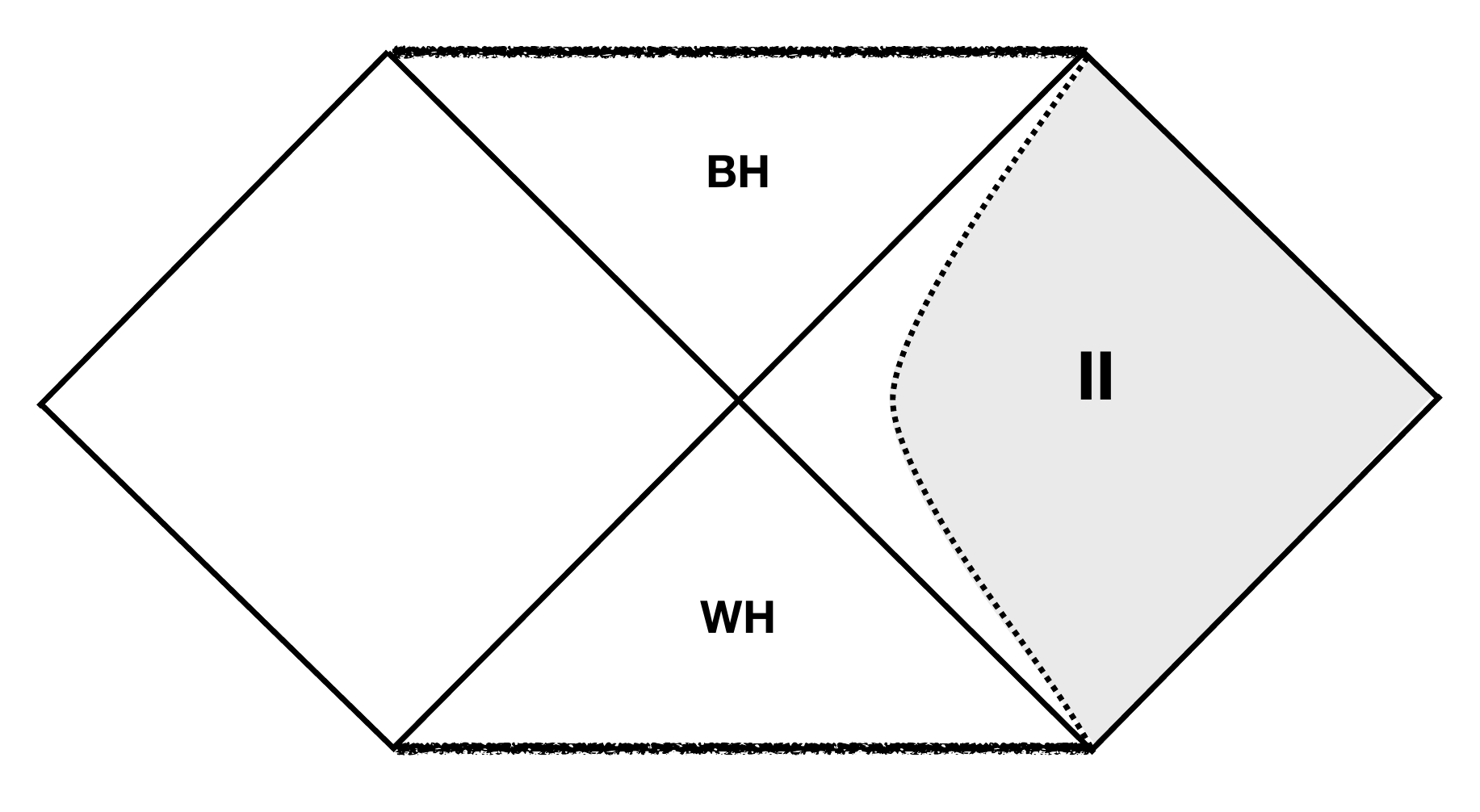}
\hfil
\includegraphics[width = .3 \textwidth]{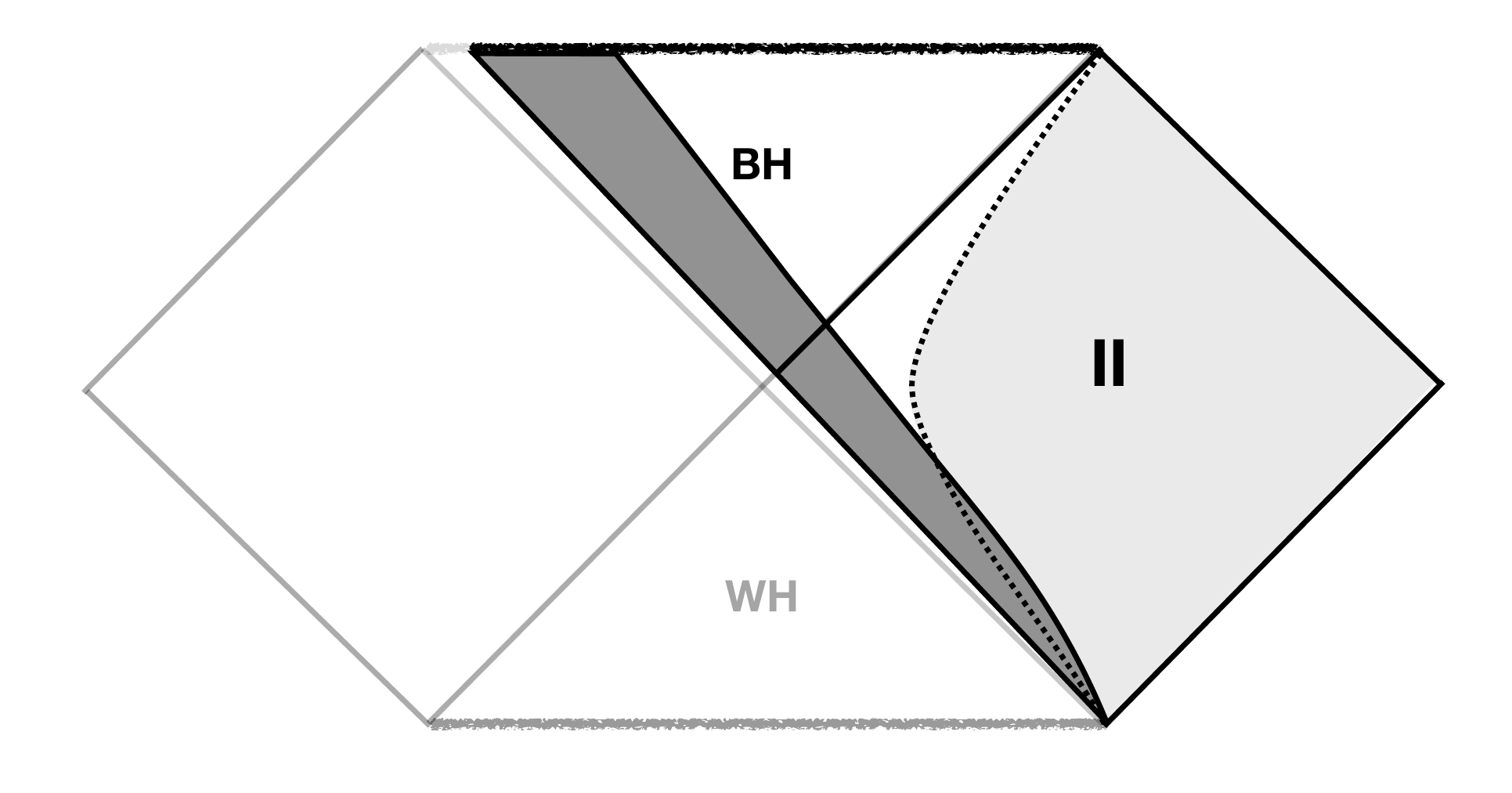}
\hfil
\includegraphics[width = .3 \textwidth]{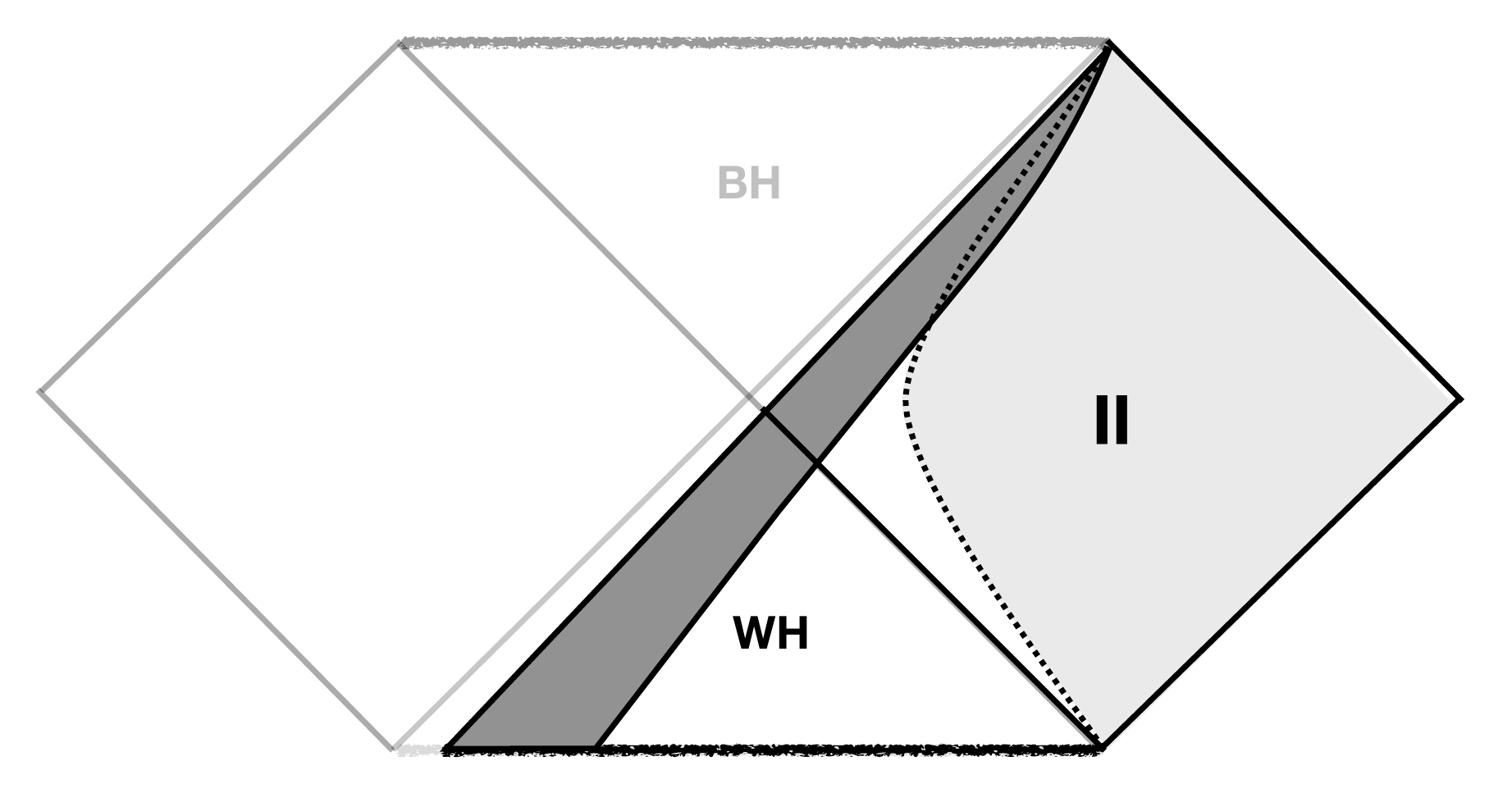}
\caption{\em {\rm Top:} in~the~extended~Schwarzschild~spacetime, which is stationary, the (light~grey) region~outside~$r=2m+\epsilon$ (dotted~line) is~equally~the outside of a~black~and~a~white hole. ~
{\rm Center:} A~collapsing~star (dark~grey) replaces~the~white~hole region~(\,{\rm WH}) in the non-stationary collapse metric.~
{\rm Bottom:} The~time~revered~process. The~difference~between~the~last~two can~only~be~detected looking~a~the~past, or~the~future.}
\end{figure}


\section{White holes}

The~difference~between~a~black~hole~and~a~white~hole is~not~very~pronounced. Observed~from~the~outside (say from~the~exterior~of~a sphere of radius $r=2m+\epsilon>2m$, where $m$~is~the~mass~of~the~hole) and~for~a~finite~amount of time, a white hole~cannot~be~distinguished~from a black hole. 

This is clear from the usual Schwarzschild line element, which is symmetric under time reversal, and therefore describes equally well the exterior of a black hole and the exterior of a white hole. Equivalently, zone II of the maximal extension of the Schwarzschild solution is equally the outside of a black hole and the outside of a white hole (see Fig.\, 1, Top). Analogous considerations hold for the Kerr solution. In other words, the continuation inside the radius $r=2m+\epsilon$ of an external \emph{stationary} black hole metric contains both a trapped region (a black hole) ad an anti-trapped region (a white hole). 
~

What distinguishes then a black hole from a white hole? 
The~objects~in the~sky~we~call `black holes' are described by~a~stationary~metric~only~approximately, and for~a~limited~time. In~their~past~(at~least)~their metric was definitely~non-stationary,~as~they~were~produced by~gravitational~collapse. 
In~this~case, the~continuation of~the~metric~inside~the~radius~$r=2m+\epsilon$~contains a trapped~region,~but~not~an~anti-trapped~region
(see Fig.\, 1,~Center). 
Viceversa,~a~white~hole~is~an~object~that is undistinguishable~from~a~black~hole~from~the
exterior and for~a~finite~time, but~in~the~future
ceases~to~be stationary and there~is~no~trapped
region~in~its~future~(see~Fig.\,1,~Bottom). 


\section{Quantum processes and time~scales}

The~classical~prediction~that~the~black~is~forever~stable is~not~reliable.~In~the~uppermost~band~of~the~central diagram~of~Fig.\,1~quantum~theory~dominates. The~death of~a~black~hole~is~therefore~a~quantum~phenomenon. The same~is~true~for~a~white~hole,~reversing~time~direction. That~is,~the~birth~of~a~white~hole~is~in~a~region where quantum~gravitational~phenomena~are~strong. 

This~consideration~eliminates~a traditional objection to 
the~physical existence of white~holes:~How~would~they~originate? 
They originate from~a~region~where~quantum phenomena dominate~the~behaviour~of~the~gravitational field. 

Such~regions~are~generated~in~particular~by~the end of~the~life~of~a~black~hole,~as~mentioned~above. Hence a~white~hole~can~in~principle~be~originated~by~a~dying black~hole. This~scenario~has~been~shown~to~be~concretely compatible with~the~exact~external~Einstein~dynamics in 
\cite{Haggard2014} 
and~has~been~explored~in
\cite{DeLorenzo2016,Christodoulou2016,Christodoulou2018,Bianchi2018,DAmbrosio2018,Rovelli2018a}.
The causal diagram~of~the~spacetime~giving~the~full~life~cycle of the black-white~hole~is~given~below~in~Fig.\,2. 

\begin{figure}[b] \label{bwh}\centering
\includegraphics[width = .25 \columnwidth]{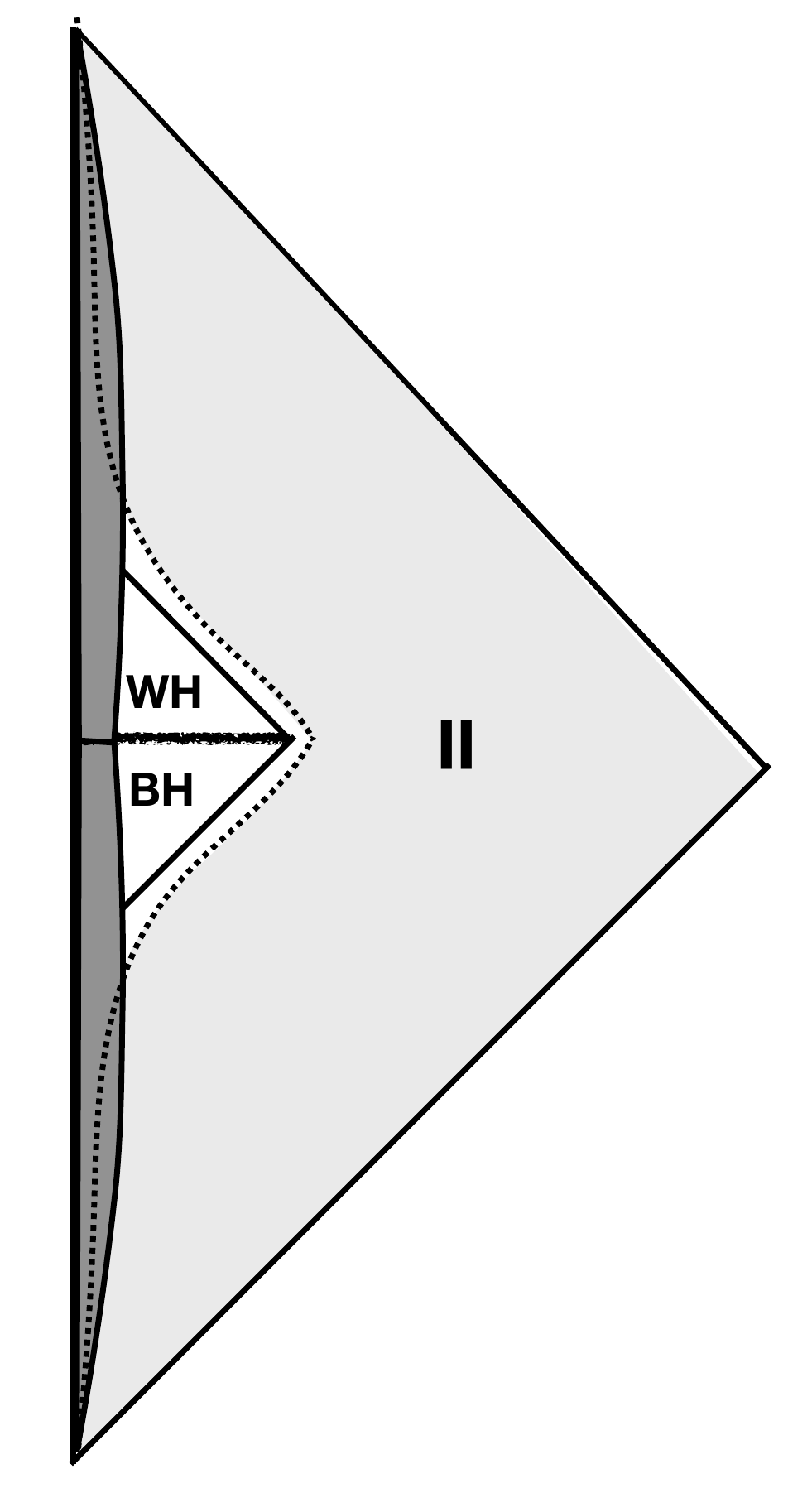}
\caption{\em The~full~life~of~a~black-white~hole.}
\end{figure}

In particular, the result of \cite{Bianchi2018} indicates that the 
black-to-white process is asymmetric in time \cite{DeLorenzo2016}
 and the time scales of the durations of the different phases are determined by the initial mass of the black hole $m_o$. 
 The lifetime $\tau_{BH}$ of the black hole is known from Hawking radiation theory to be at most of the order
\begin{equation}
\tau_{BH}\sim m_o^3
\end{equation}
in~Planck~units~$\hbar=G=c=1$. 
This~time~can~be~as shorter~as~$\tau_{BH}\sim~m_o^2$~because~of~quantum~gravitational effects \cite{Rovelli2014h,Haggard2014,DeLorenzo2016,Christodoulou2016,Christodoulou2018} (see~also~\cite{Gregory:1993vy,Casadio:2000py,Casadio:2001dc,Emparan:2002jp,Kol:2004pn})
but~we~disregard~this possibility~here.
The~lifetime~$\tau_{W\!H}$~of~the~white~hole~phase is~longer~\cite{Bianchi2018}:
\begin{equation}
      \tau_{WH}\sim m_o^4 
\end{equation}
in~Planck~units.
~That~is,~in~arbitrary~units:
\begin{equation}
      \tau_{WH}= \frac{m_o}{m_{Pl}} \tau_{BH}\,, 
\end{equation}
where  $m_{Pl}$ is the Planck mass.  
The tunnelling process itself from black to white takes a time of the order of the current mass at transition time \cite{Christodoulou2018}. 
The area of the horizon of the black hole decreases with time because of Hawking evaporation, decreasing from $m_o$ to the Planck mass $m_{Pl}$. 
At this point the transition happens and a white hole of mass of the order of the Planck mass is formed. 


\section{Timescales}

Consider the hypothesis that white-hole remnants are a constituent of dark matter. 
To give an idea of the density of these objects, a local dark matter density of the order of $0.01 M_\odot/pc^3$ corresponds to approximately one Planck-scale remnant, with the weight of half a inch of human hair, per each $10.000 Km^3$. 
For these objects to be still present now we need that their lifetime be larger or equal than the Hubble time $T_H$, that is
\begin{equation}
      m_o^4\ge T_H. 
\end{equation}
On the other hand, since the possibility of many larger back holes is constrained by observation, we expect remnants to be produced by already evaporated black holes, therefore the lifetime of the black hole must be shorter than the Hubble time. 
Therefore 
\begin{equation}
      m_o^3 < T_H. 
\end{equation}
This gives an estimate on the possible value of $m_0$:
\begin{equation}
      10^{10} gr \le m_o^3 < 10^{15} gr.
\end{equation}
These are the masses of primordial black holes that could have given origin to dark matter present today in the form of remnants. 
Their Schwarzschild radius is in the range 
\begin{equation}
      10^{-18}~cm \le R_o < 10^{-13}~cm\,.
\end{equation} 
According to a commonly considered theory of primordial black-hole formation, black holes of a given mass could have formed when their Schwarzschild radius was of the order of the cosmological horizon. 
Remarkably, the horizon was presumably in the above range at the end of inflation, during or just after reheating. Which happens to be precisely the epoch where we expect primordial black hole formation, namely shortly after reheating.   
This concordance supports the plausibility of the proposed scenario; 
that is, if the lifetimes in the model we are considering are correct, the black holes formed in that period are around us as remnants: 
they have have already ended the Hawking evaporation but the resulting white holes have not had the time to dissipate yet.


\section{Stability}

Large~classical~white~holes~are~unstable 
(see for instance Chapter 15 in \cite{Frolov2012} and references therein). 
The reason can be understood as follows. 
The spacetime depicted in the Center panel of Fig.\,1 does not change much under a small arbitrary modification of its initial conditions on past null infinity; but it is drastically modified if we modify its final conditions on future null infinity. 
This is intuitively simple to grasp: if we sit on future null infinity and look back towards the hole, we see a black disk. 
This is the final condition. 
A slightly perturbed final condition includes the possibility of seeing radiation arriving from this disk. 
This is impossible in the spacetime of the Center panel of Fig.\,1, because of the huge red shift of the radiation moving next to the horizon, but it is possible in the Top panel spacetime, because the radiation may have crossed over from the other asymptotic region. 

The same is true for a white hole, reversing the time direction. In the spacetime depicted in the Bottom panel, with some radiation, there is necessarily a dark spot in the \emph{incoming} radiation from past null infinity. If we perturb this configuration, and add some incoming radiation in this dark spot, the evolution generically gives the spacetime of the Top panel. Physically, what happens is that this radiation moves along the horizon, is blue shifted, can meet radiation coming out of the white hole and this is more mass that $m$ at a radius $2m$: it is mass inside its Schwarzschild radius. At this point the region is trapped, and a black hole forms. Consequently the evolution of the perturbed initial conditions yields the spacetime on the Top, not the one on the Bottom: 
the white hole is unstable and decays into a black hole. 

This is the standard `instability of white holes'. 
How does this instability affect the remnants formed at the end of a black hole evaporation? ~
The wavelength of the perturbation needed to trigger the instability must be smaller that the the size of the hole \cite{Frolov2012}. 
It was observed in \cite{Bianchi2018} that to trigger the instability of a Planck size white hole we need trans-Planckian radiation, and this is likely not be allowed by quantum gravity. Below we explore the issue in more detail building a quantum model to describe the processes involving black and white holes. 
\section{Black and white hole processes}

\begin{figure}[b]
\includegraphics[width = .3 \textwidth]{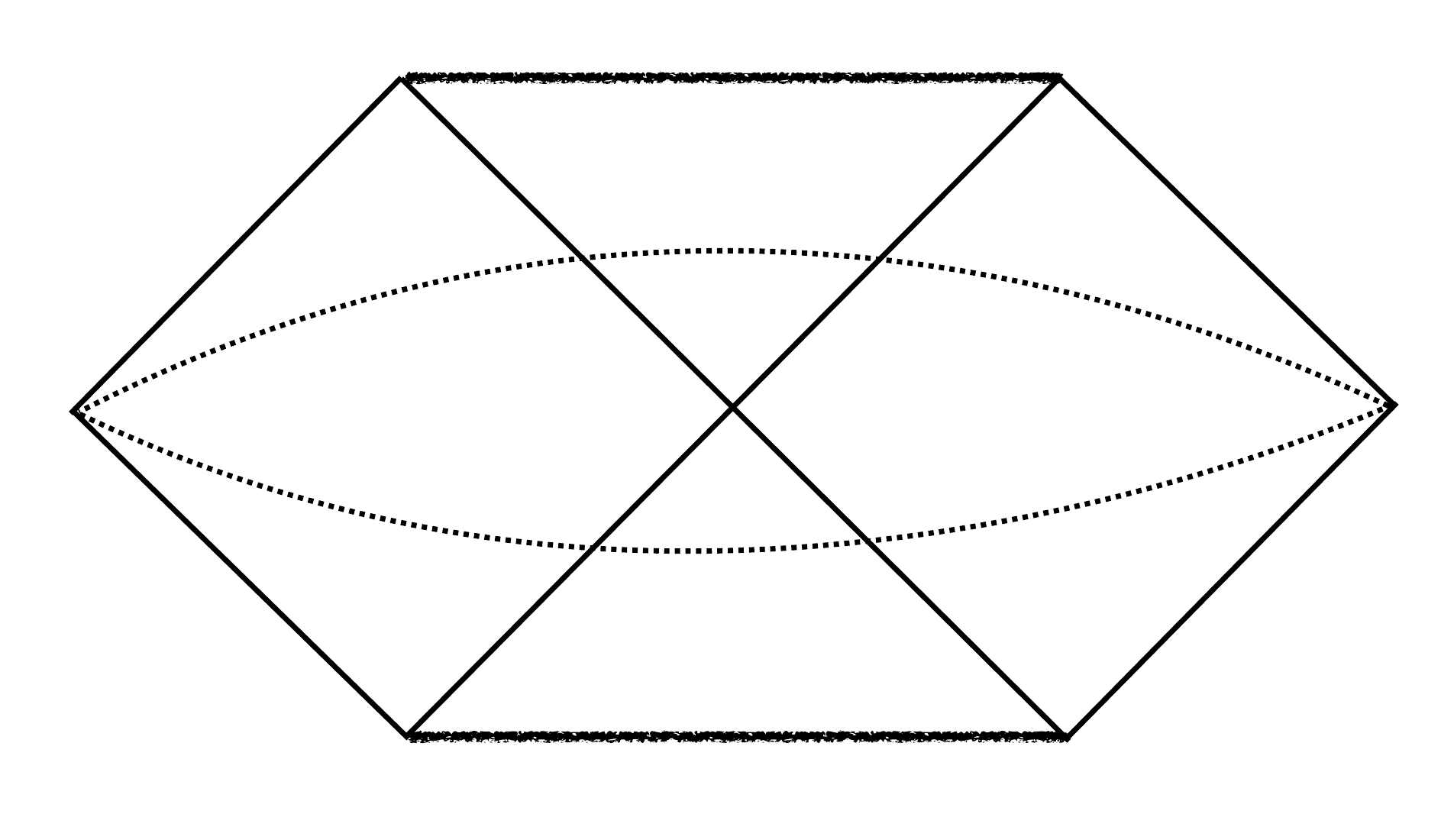}
\hfil
\includegraphics[width = .3 \textwidth]{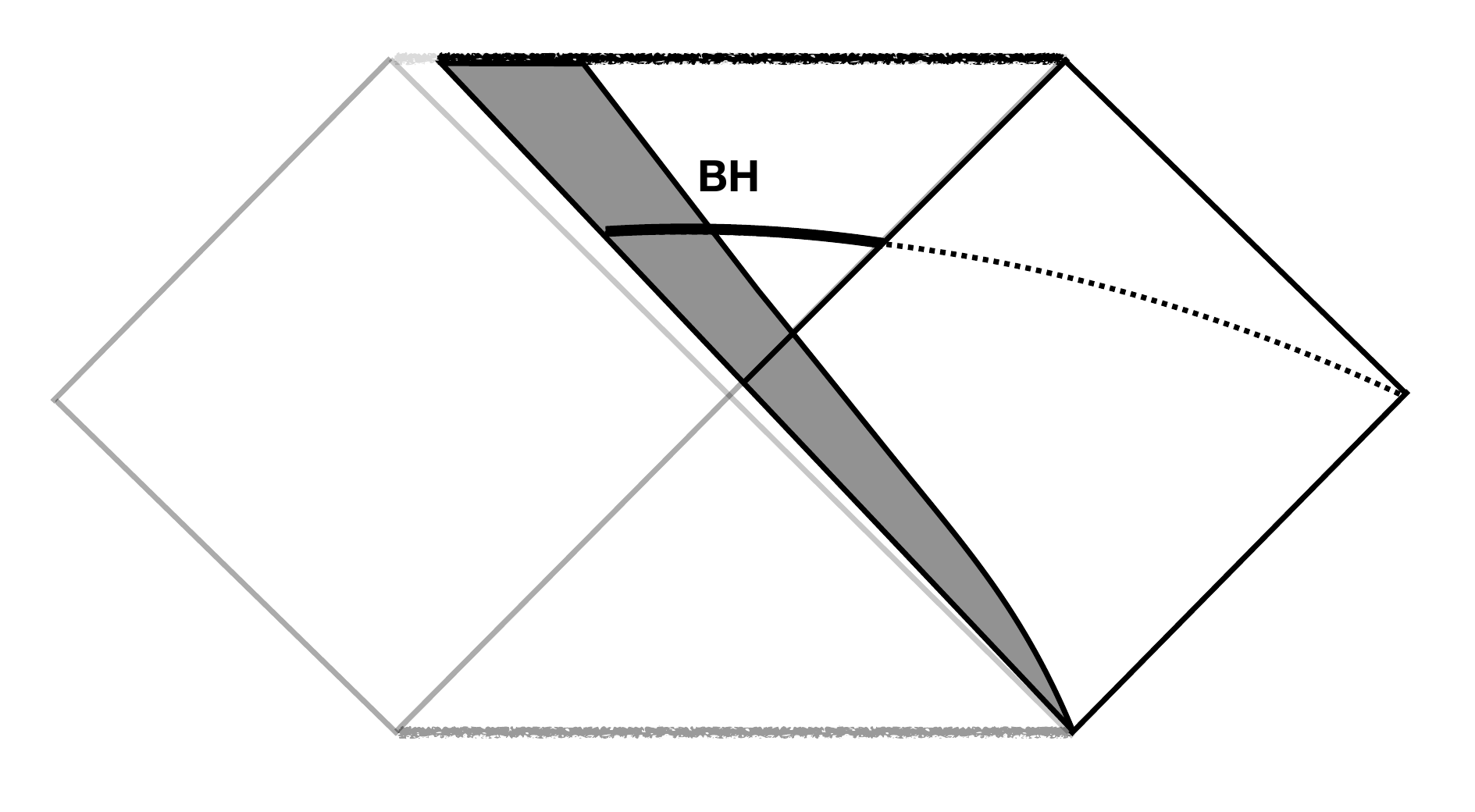}
\hfil
\includegraphics[width = .3 \textwidth]{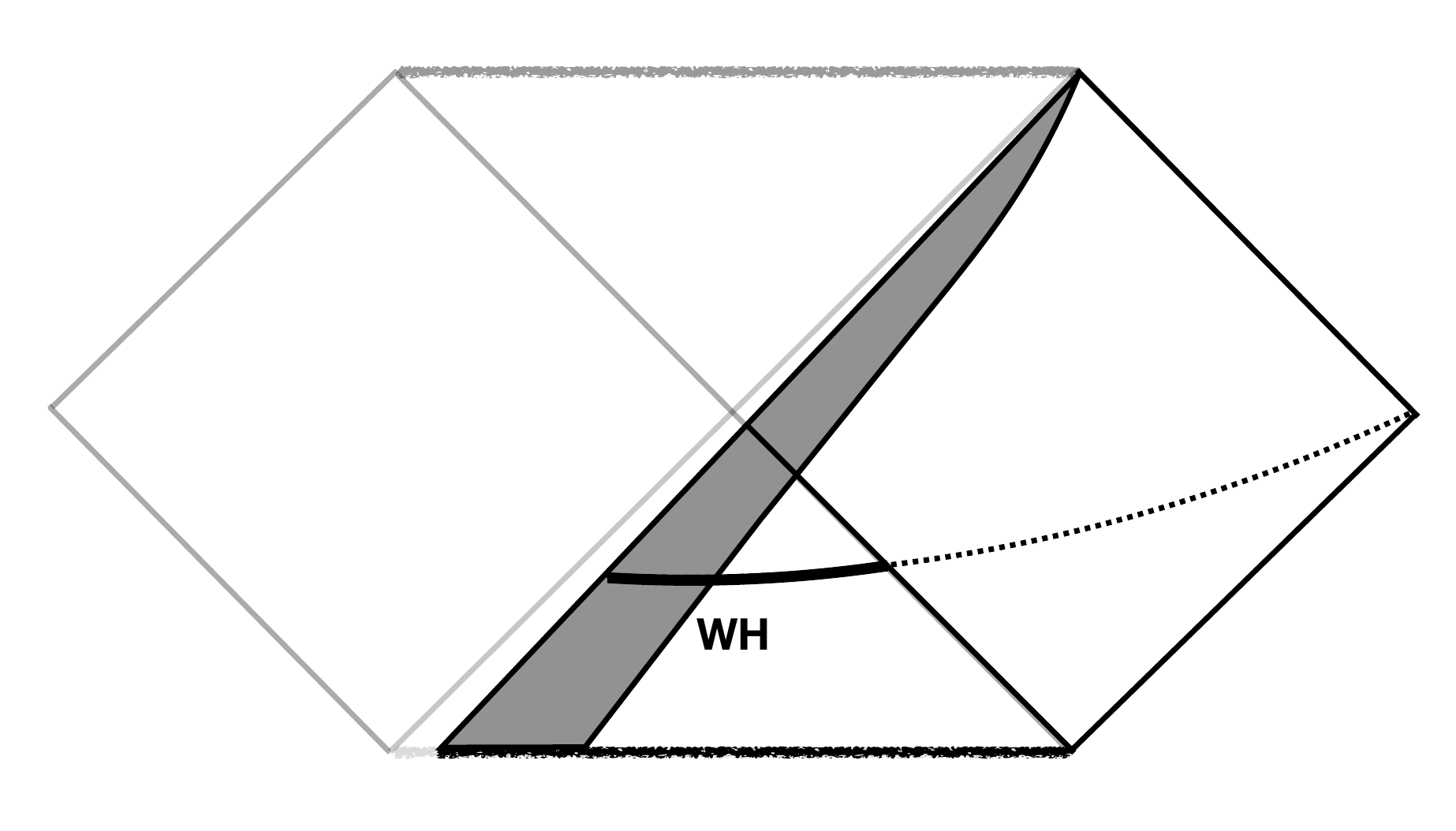}
\caption{\em {\rm Top:} Cauchy surfaces in extended Schwarzschild spacetime below and above the central sphere.
~
{\rm Center:} Internal portion of a Cauchy surface describing a black hole formed by a collapsed star.
~
{\rm Bottom:} Its time reversal, or white hole.}
\label{cauchy}
\end{figure}

Consider a (spherically symmetric) Cauchy surface $\Sigma$ in an extended Schwarzschild spacetime with mass $m$. $\Sigma$ can cut the horizon below or above the central sphere (the bifurcating horizon). 
See the Top panel of Fig.\,\ref{cauchy}. 
If above, we say that $\Sigma$ contains a black hole; if below, we say it contains a white hole. In either case, the Top asymptotic region and a portion of the interior of the hole can be replaced by a finite matter-filled interior. 
In this case the interior portion of $\Sigma$ has a finite volume $v$. See Fig.\,\ref{cauchy}. One possibility is to fix this interior portion of $\Sigma$ to have constant trace of the exterior curvature, or, equivalently, maximal volume \cite{Christodoulou2015,Bengtsson2015,Ong2015,Christodoulou2016a}. Then the surface is determined by its intersection with the horizon (other interesting gauge fixing are possible).

Let $|H,m,v\rangle$ where $H=B,W$ (for `Black' and `White') denote a coherent (semiclassical) quantum state of matter and geometry on the portion of $\Sigma$ that lies inside the horizon, corresponding respectively to the Center and Bottom panels of Fig.\,\ref{cauchy}, namely entering respectively the trapped ($H=B$) or anti-trapped ($H=W$) regions. Here $m$ and $v$ are the (expectation values of) mass and interior volume of the Schwarzschild geometry. We take here an approximation where these are the only relevant degrees freedom. We assume black hole states $|B,m,v\rangle$ and white hole states $|W,m,v\rangle$ to be orthogonal states in the common Hilbert space $\tilde{\cal H}$ of the quantum states of geometry and matter inside a sphere, of Schwarzschild radius $r=2m$. This is a reduced model since we disregard internal degrees of freedom others than $v$.
We are interested in the evolution of the state as the surface $\Sigma$ moves up in time.\\

Let's label the position of $\Sigma$ with a temporal parameter {t}. For a black hole, it is natural to identify $t$ with the advanced time {v} and for a white hole, it is natural to identify it with the retarded time {-u}. So let's define
\begin{equation}
d{\rm t}=d{\rm v},\ {\rm for}\ H=B, \ \ {\rm and}\ \ d{\rm t}=-d{\rm u}, \ {\rm for}\ H=W,
\end{equation}
with an arbitrary origin for the {t} label.
%
A number of processes can occur as the surface $\Sigma$ moves up in time.  
We list them here using relativistic units $G=c=1$ and keeping $\hbar$ explicit to distinguish classical from quantum phenomena. 
\begin{enumerate}
\item {\em Black hole volume increase and white hole volume decrease}
\begin{eqnarray}
|B,m,v\rangle&\to& |B,m,v+\delta v\rangle, \\
|W,m,v\rangle&\to& |W,m,v-\delta v\rangle.
\end{eqnarray}
This is simply determined by the Einstein's equations if nothing else happens. The variation is computed in \cite{Christodoulou2015} to be governed by 
\begin{equation}
\frac{dv}{d{\rm t}}=\pm3\sqrt3 \pi m_o^2. 
\end{equation}
where $m_o$ is the initial mass of the black hole and the sign is plus for a black hole and minus for white hole. 
\item {\em White to black instability}
\begin{equation}
|W,m,v\rangle\to |B,m,v\rangle.
\end{equation}
This process is allowed by classical general relativity in the absence of any perturbation when there is a second asymptotic region, as it is apparent from the Top panel of Fig.\,\ref{cauchy}; but it can also be triggered by an external perturbation \cite{Frolov2012}. Notice that the volume does not change: this is due to the fact that this is a local process in the horizon region, which does not modify the interior. The lifetime of a white hole under decay to a black hole has been estimated to be proportional to its Schwarzschild radius \cite{Frolov2012}:
\begin{equation}
\tau_{\,W\to B}\sim m.
\end{equation}
This is equivalent to a transition probability per unit of time
\begin{equation}
p \sim m^{-1}. 
\end{equation}
\item {\em Hawking evaporation}
\begin{equation}
|B,m,v\rangle\to |B,m-\delta m,v\rangle.
\end{equation}
This is a process that decreases the mass of a black hole, produced by negative energy entering the hole when a Hawking quantum is radiated. It is a phenomenon described by the classical backreaction on the geometry of the dynamics of a quantum field. Hawking radiation theory gives
\begin{equation}
\frac{dm}{d{\rm t}} = \frac{\hbar}{m^2}.
\end{equation}
Giving the lifetime for a massive black hole 
\begin{equation}
\tau_{B}\sim \frac{m^3}{\hbar}.
\end{equation}
\item {\em Black to white tunnelling}
\begin{equation}
|B,m,v\rangle\to |W,m,v\rangle. 
\end{equation}
This is a genuine quantum gravitational process \cite{Rovelli2014b,Rovelli2014,Bianchi2018}. Its probability per unit of time is still unclear. We take here the conservative estimate derived in \cite{Christodoulou2018} using covariant Loop Quantum Gravity \cite{Rovelli:2014ssa}, which agrees with the semiclassical expectation for tunnelling phenomena, namely that this probability is suppressed by the semiclassical standard tunnelling factor
\begin{equation}
e^{-\frac{S}{\hbar}} \sim e^{-\frac{m^2}{\hbar}}
\end{equation}
where $S$ is a typical action for the transition. On dimensional grounds, this suggests a tunnelling probability per unit time
\begin{equation}
p \sim e^{-\frac{m^2}{\hbar}}/m
\end{equation}
Here we have assumed for simplicity that the internal volume $v$ is conserved in this transition. A more precise account of this process will be studied elsewhere (for the tentative phenomenology derived from this process, see \cite{Barrau2014e,Barrau2014yka,Barrau2015uca,Barrau2016,Vidotto:2016jqx,Barrau2018a,Vidotto2018}). 
\end{enumerate}

\section{Dynamical evolution}

The ensemble of the processes listed above can be described as an evolution in {t} 
\begin{equation}
i\hbar\,\partial_{\rm t} |\psi\rangle=H |\psi\rangle
\end{equation}
for a two component state
\begin{equation}
|\psi\rangle=\left(
\begin{array}{l}
\,B(m, v) \\[1mm]
W(m, v)
\end{array}
\right)
\end{equation}
governed by the Hamiltonian 
\begin{equation}
H=\left(
\begin{array}{cc}
\!m+3\sqrt3 \ i \pi m_o^2 \, \frac{\partial}{\partial v}-i\,\frac{\hbar^2}{m^2}\frac{\partial}{\partial m}\! & b\frac{\hbar}{m} 
\\[1em]
c \frac{\hbar}{m} e^{-m^2/\hbar} &\!m-3\sqrt3 \ i \pi m_o^2 \, \frac{\partial}{\partial v}
\end{array}
\right)
\label{H}
\end{equation}
where we have added also a diagonal energy term proportional to the mass in order to obtain the standard energy phase evolution, and $c$ and $b$ are constants of order unit.

We now ask what are the stable or semi-stable states of the hole {\em as seen from the exterior}. 

A macroscopic black hole with mass $m$ much larger than the Planck mass $m_P=\sqrt{\hbar}$ is stable when seen from the exterior for a (long) time span of the order $m^3/\hbar$, which is the Hawking evaporation time. The stability is due to the fact that process (1) does not affect the exterior, process (2) does not concern black holes and process (4) is strongly suppressed for macroscopic holes. 

A macroscopic white hole, on the other hand, is not so stable, because of the fast instability of process (2). As basic physics is invariant under time reversal, one may wonder what breaks time reversal invariance here. What breaks time reversal invariance is the notion of stability that we are using. This is a stability under small fluctuations of the \emph{past} boundary conditions. If instead we asked about stability under small fluctuations of the \emph{future} boundary conditions, we would obviously obtain the opposite result: macroscopic white holes would be stable while macroscopic black holes would not. 

The question we are interested in is what happens (generically) to a large macroscopic black hole if it is not fed by incoming mass. Then two processes are in place: its Hawking evaporation for a time $\sim m^3/\hbar$ (process 3) and the internal growth of $v$ (process 1). This continues until process (4) becomes relevant, which happens when the mass is reduced to order of Planck mass. At this point the black hole has a probability of order one to tunnel into a white hole under process (4). But a white hole in unstable under process (2), giving it a finite probability of returning back to a black hole. Both processes (4) and (2) are fast at this point. Notice that this happens at large $v$, therefore in a configuration that classically is very distant from flat space, even if the overall mass involved is small. 

As energy is constantly radiated away and no energy is fed into the system, the system evolves towards low $m$. But $m$ cannot vanish, because of the presence of the interior: in the classical theory, a geometry with larger $v$ and small $m$ is not contiguous to a Minkowski geometry, even if the mass is small. Therefore in the large $v$ region we have $m>0$. Alternatively, this can be seen as a hypothesis ruling out macroscopic topology change. 

But $m$ cannot be arbitrarily small either, because of quantum gravity. The quantity $m$ is defined locally by the area of the horizon \,$A=16\pi G^2\,m^2$\, and $A$ is quantized. According to Loop Quantum Gravity \cite{Rovelli:2010bf} the eigenvalues of the area of {\em any} surface are \cite{Rovelli1994a} 
\begin{equation}
A= 8\pi \,\hbar G\, \sqrt{j(j+1)}
\end{equation}
where we have taken the Immirzi parameter to be unit for simplicity. The minimal non-vanishing eigenvalue is 
\begin{equation}
a_o= 4\sqrt{3} \pi \,\hbar G
\end{equation}
and is called the `area gap' in loop quantum cosmology \cite{Ashtekar2015}. This gives a minimal non-vanishing mass $\mu$ defined by $a_o=16\pi G^2 \mu^2$, that is
\begin{equation}
\mu\equiv \frac{3^{\frac14}}2\sqrt\frac{\hbar}{G}.
\end{equation}
(we have momentarily restored $G\ne 1$ for clarity.) Radiating energy away brings down the system to the $m=\mu$ eigenspace. Consider now states that are \emph{eigenstates} of $m$ with the minimal value $m=\mu$ and denote them $|B,\mu,v\rangle$ and $|W,\mu,v\rangle$. The dynamics governed by the above Hamiltonian allows transition between black and white components. This is a typical quantum mechanical situation where two states, here $|B,\mu,v\rangle$ and $|W,\mu,v\rangle$, can dynamically turn into one another. Let us we disregard for a moment $v$, which is invisible from the exterior, and project $\tilde{\cal H}$ down to a smaller state space ${\cal H}$ with basis states $|H,\mu\rangle$. This is a two dimensional Hilbert space with basis vectors $|B,\mu\rangle$ and $|W,\mu\rangle$. Seen from the exterior, the state of $\Sigma$ will converge to ${\cal H}_\mu$. 

The Hamiltonian acting on this subspace is
\begin{equation}
H=\left      (
\begin{array}{cc}
\mu & \frac{b\hbar}\mu \\ 
\frac{a\hbar}\mu & \mu
\end{array}
\right )
\label{H}
\end{equation}
where $a=c e^{-\frac{\sqrt{3}}4}$. Quantum mechanics indicates that in a situation where the system can radiate energy away and there are possible transitions between these two states, the actual state will converge to a quantum state which is a quantum superposition of the two given by the lowest eigenstate of $H$. This is 
\begin{equation}
|R\rangle = \frac{\sqrt{\frac{a}{b}} |B,\mu\rangle - |W,\mu\rangle}{\sqrt{1+\frac{a}{b}}}
\end{equation} 
($R$ for `Remnant') and has eigenvalue $\mu-\hbar\sqrt{ab}/\mu$. If the amplitude $b$ of going from black to white is larger than the amplitude $a$ of going from white to black (as it seems plausible), the state is dominated by the white hole component. A related picture was been considered in \cite{Barcelo:2015uff,Barcelo2016,Garay2017}: a classical oscillation between black and white hole states. 

In a fully stationary situation, the mass $m$ is equal to the Bondi mass, which generates time translations at large distance from the hole in the frame determined by the hole. (Quantum gravity is locally Lorentz invariant \cite{Rovelli2003,Rovelli:2002vp} and has no preferred time \cite{Rovelli2011h} but a black hole in a large nearly-flat region determines a preferred frame and a preferred time variable.) Keeping possible transitions into account there is a subtle difference between the mass $m$, determined locally by the horizon area, and the energy of the system, which is determined by the full Hamiltonian that includes the interaction terms. Being an eigenstate of the energy, $|R\rangle $ is a stationary state, as far as the external dynamics is concerned, and as long as the internal volume remains large. This is the stable remnant that forms after the end of Hawking evaporation. 

On a time scale of order $m_o^4$, on the other hand, the internal volume of the white hole component of this state can shrink to zero, and a transition to Minkowski space becomes probable, a process not explicitly contemplated by the simple model given. 

\section{Conclusions}

We have computed the formation time of primordial black holes that could have given rise to remnants forming a component of dark matter today, and we have found that this formation time sits at a cosmological epoch after the end of inflation, compatible with the current observation of dark matter.

We have addressed the issue of remnant stability by building a simple quantum model that takes white hole instability explicitly into account, and shown that its only consequence is to induce a quantum superposition between Planckian-area quantum white and black hole states.

This is a preliminary crude quantum model. It disregards the detailed dynamics around the minimal radius \cite{DAmbrosio2018} and in the tunnelling region \cite{Christodoulou2018,Rovelli2018a} and, importantly, the internal dynamics generated by the in-falling component by the Hawking's radiation. The consistency of this picture depends on the hypothesis that there is no macroscopic topology change and a large black hole interior does not just magically disappear into nothing. 

Still, its preliminary indications support the possibility of stable remnants, make pretty clear that white hole instability is not an issue for their existence, and therefore remnants may well be components of the observed dark matter. 

An alternative possibility, where dark matter is formed by remnants from a pre-big-bang phase in a bouncing cosmology, is explored in a companion paper \cite{Rovelli:2018ee}. 

\vskip3em
{\bf Acknowledgements} ~
Thanks to Alejandro Perez, Simone Speziale, Tommaso De Lorenzo, Giorgio Sarno, Eugenio Bianchi, Hal Haggard for helpful exchanges. \\
The work of FV at UPV/EHU is supported by the grant IT956-16 of the Basque Government and by the grant FIS2017-85076-P (MINECO/AEI/FEDER, UE).

\vfill


\providecommand{\href}[2]{#2}\begingroup\raggedright\endgroup

\end{document}